%% file: barger.tex
\documentstyle[12pt,aaspp4,overcite,flushrt,naturestyle]{article}

\include{basic}

\def\tc4{\tau({\rm C~IV)}}

\def\ten#1{\times 10^{#1}}

\begin{document}

\title{Dusty star forming galaxies at high redshift}

\author{A. J. Barger$^{\ast}$, L. L. Cowie$^{\ast}$, 
D. B. Sanders$^{\ast}$, E. Fulton$^{\ast}$, Y. Taniguchi$^{\dag}$,
Y. Sato$^{\dag}$, K. Kawara$^{\ddag}$, \& H. Okuda$^{\S}$}

\leftline{$^{\ast}$\ Institute for Astronomy, University of Hawaii, 2680
Woodlawn Drive, Honolulu, HI 96822}

\leftline{$^{\dag}$\ Astronomical Institute, Tohoku University, Aoba, 
Sendai 980-77, Japan}

\leftline{$^{\ddag}$\ Institute of Astronomy, University of Tokyo, 
2-21-1 Osawa, Mitaka, Tokyo 181, Japan}

\leftline{$^{\S}$\ Institute of Space and Astronautical Science,
Yoshinodai, Sagamihara, Kanagawa 229, Japan}

%\centerline{Accepted for publication in {\it Nature}}

\noindent
\rule{6.5in}{0.5mm}

\vspace*{0.05in}
\noindent
{\bf 
The global star formation rate in high redshift galaxies, based on optical
surveys, shows a strong peak at a redshift of $z\sim 1.5$, which implies that we
have already seen most of the formation. High redshift galaxies may, however,
emit most of their energy at submillimeter wavelengths if they contain
substantial amounts of dust. The dust would absorb the starlight and reradiate
it as far-infrared light, which would be redshifted to the
submillimeter range. Here we report a deep survey of two blank regions of sky
performed at submillimeter wavelengths (450 and 850$\micron$). If the sources we
detect in the 850$\micron$ band are powered by star formation, then each must be
converting more than 100 solar masses of gas per year into stars, which is
larger than the maximum star formation rates inferred for most
optically-selected galaxies. The total amount of high redshift star
formation is essentially fixed by the level of background light, but
where the peak occurs in redshift for the submillimeter is not yet
established. However, the background light contribution from only the 
sources detected at 850$\micron$ is already comparable to that from
the optically-selected sources. Establishing the main epoch of star 
formation will therefore require a combination of optical and submillimeter
studies.
}

In recent years high redshift optical galaxy searches have become increasingly
successful at uncovering significant populations of galaxies that are likely
to be in early phases of evolution. 
However,
the global star formation rate (SFR) inferred\cite{madau,steidel,cowiehu} 
omits the many fainter sources that are now being detected \cite{cowie2,hu1}.
Furthermore, the effects of dust can cause
the SFRs in the detected UV-bright objects to be grossly underestimated
(see, e.g., ref.~\citen{heck}), and many rapid star forming galaxies may 
even be omitted from the optical samples.  

Nearby star forming galaxies
emit a large fraction of their bolometric luminosity in the far infrared
waveband, which for distant sources is redshifted into the submillimeter
waveband. Because the spectra of these star forming galaxies are very steep,
if they are at large redshifts their flux density
decreases much less rapidly with increasing redshift than is expected
according to the inverse square law.  Thus, galaxies at high redshifts
are likely to make a substantial contribution to the submillimeter counts
\cite{blainlongair}.
In a pioneering paper \cite{smail}, Smail et al.\ reported the detections of 
submillimeter sources in the fields of two rich clusters of galaxies
using the new SCUBA bolometer
array \cite{holland} on the 15-meter James Clerk Maxwell Telescope.
These clusters were selected because they are strong gravitational lenses 
which magnify any submillimeter sources lying behind them.  The objects
in the A370 field include one spectacular source,
SMM~02399$-$136, with an 850$\micron$ flux of 26~mJy
($1\ \rm{Jy} = 1.0\times 10^{-23}\ 
{\rm erg}\ {\rm s}^{-1}\ {\rm cm}^{-2}\ {\rm Hz}^{-1}$). This source has 
subsequently been shown \cite{ivison} to be
an AGN embedded in a starburst at $z = 2.8$. 
Gravitational lensing greatly improves source detectability but entails
an uncertain correction to remove the estimated magnification of the
sources. The
most likely amplification factor for the above source was 
given\cite{ivison} as 2.5 with a firm upper limit of 5.
The magnification problem can be avoided if blank field observations are
made with longer exposure times. That
is the nature of the experiment reported here and in other
upcoming blank field survey papers by UK and UK-Canadian teams.

We obtained SCUBA observations of two blank field areas, one in the 
Lockman hole and one in the Hawaii deep field region SSA13.
Our data were taken during 8 eight-hour shifts in March 1998 and
8 eight-hour shifts in May 1998.
Fully sampled wavelength maps were completed simultaneously in
both the long (850$\micron$) and short (450$\micron$) wavelength arrays,
but the sky at 450$\micron$ is considerably more opaque.
Although two of the 16 shifts were too poor to contribute significantly 
to the observations, much of the run had superb weather conditions and 
produced sensitive maps at both wavelengths.  For the 850$\micron$ 
maps, which are of primary interest, the total integration time was 
39 hours on the Lockman hole region and 51 hours on SSA13. 

At 450$\micron$ no sources were detected at the $3\sigma$\ level 
of $25$\ mJy in either
field.  At 850$\micron$ two sources having fluxes $>3$\ mJy were
detected at the $>3\sigma$\ level, one in each field; these are listed
in Table~1.  The 850$\micron$ maps are described and shown in Figure~1. 
The absolute value of the noise was found from the dispersion of the points in
the map, excluding the regions around the brightest sources (see Table~1).  
Based on predictions of source confusion\cite{blain1}, 
the instrumental and sky noise are expected
to dominate the total noise.
The completeness of source recovery as a function of flux was tested
using Monte Carlo simulations. Five sources at each of a range of flux 
levels were
simulated using the beam derived from the calibration source and were
added into the field. The source was considered to be recovered if
it was detected at the $3\sigma$\ level by the source detection algorithm.
This procedure was repeated a large number of times at each flux.
Recovery of sources brighter than 3\ mJy
was essentially complete, except for a small number of cases where the
objects overlapped.  
The fall off in recovery was extremely rapid, with 80 per cent of the sources
recovered around 3\ mJy and only 20 per cent by 2\ mJy.
Inspection of the negative images yielded no sources
with fluxes greater than 3\ mJy.
Given that we detect only two sources in our data maps with fluxes
above 3\ mJy using the same $3\sigma$ threshold, we may conclude that the
surface density of sources above 3~mJy is 800 (290--1900)\ ${\rm deg}^{-2}$,
where the numbers in parentheses are the $\pm 1\sigma$ range.  
Our number is smaller than the tentative estimate of $2500\pm 1400$
deg$^{-2}$ above 4\ mJy made in ref.~\citen{smail} on the basis of
observations of cluster-lensed sources.  This difference may
partially reflect the uncertainty in the magnification 
correction but could also arise in large part from the small-number 
statistics of the observations. Our value
corresponds to a $\nu\,S_{\nu}$, where $S_{\nu}$ is the sky surface 
brightness, of $3.7\ten{-8}~{\rm ergs\ cm}^{-2}\ s^{-1}\ {\rm ster}^{-1}$,
which is a factor of a few less than current limits on the
extragalactic background light\cite{puget,burigana,fixsen,schlegel,hauser}.

Our strongest source in the two fields is the $4.6$\ mJy
($\sim 6\sigma$) source in the Lockman hole.
In Figure~2 we show a deep $K'$ image of the Lockman hole SCUBA field 
with a 3\,mJy/beam contour superimposed on the bright source.
We also show deep zoomed $K'$ and 
$B$-band images of the region surrounding the source.
An optical spectrum obtained with the Low Resolution Imaging Spectrometer
(LRIS\cite{oke}) on the Keck 10-meter telescope 
reveals only continuum emission.
Since [OII] 3727\AA\ would shift out of the observed spectral range for
$z\approxgt 1.5$, this suggests that the source lies above this redshift 
limit.  
However, the presence of the object in the $B$-band image places an upper 
limit on the redshift of the source of $z\approxlt 3.5$.

Given these redshift constraints, the only class of local objects whose 
spectral 
energy distribution (SED) could provide an approximate match to the $K'$ and 
850$\micron$ flux measurements of our brightest source is an ultraluminous
infrared galaxy \cite{sanders}.
For the following discussion we assume that the 
SED of the ``prototypical'' ultraluminous infrared galaxy 
Arp~220 is appropriate for our object.  The spectral shape of
Arp~220 can be represented \cite{klaas} by a modified blackbody with 
emissivity $\lambda^{-1}$ and a dust temperature of 
$T=47$\ K. If we place an Arp~220 source at a redshift $z$, the
effective blackbody temperature becomes $T/(1+z)$ due to the expansion
of the Universe. The far infrared luminosity ($L_{\rm FIR}$) can then
be related to the observed 850$\micron$ flux of our brightest source. 
Assuming the Arp~220 temperature of $T=47$\ K,
we find $L_{\rm FIR}=(0.9 - 1.6)\ten{12}\lsun\,h_{75}^{-2}$\
for $\qno = 0.5$ and 
$(1.2 - 6.1)\ten{12}\lsun\,h_{75}^{-2}$\ for $\qno = 0.02$
over the redshift range $0.5 < z < 3.5$.
Submillimeter observations of nearly all luminous infrared galaxies have been
reasonably fit by single temperature dust models\cite{sanders} with
$T=30-50$\ K; thus, the temperature dependence of $L_{\rm FIR}$
introduces a factor of only about 0.2--1.2 uncertainty in the above 
$L_{\rm FIR}$ numbers.

Although we do not have radio spectral indices or diagnostic spectral
line ratios that are traditionally used to discriminate between thermal 
and non-thermal energy sources, it is probable, as 
in the low-$z$\ ultraluminous infrared galaxies, that the energy in
our submillimeter sources is powered by a mixture of AGN and star formation
with a substantial fraction arising from the star formation.
If we assume that all of the energy originally in
the rest frame ultraviolet is reradiated into the FIR and take the
extreme case of zero contribution to the FIR emission from an obscured
AGN, then the FIR 
luminosity provides a measure of the current SFR of massive 
stars\cite{scoville,thronson}
${\rm SFR} = \Psi\ 10^{-10}\ (L_{\rm FIR}/\lsun)\ \msun\ {\rm yr}^{-1}$,
where $\Psi=0.8-2.1$.
To produce all of the luminous energy of our brightest source from intense 
ongoing star formation would therefore require star formation rates of 
$(140 - 250)\msun\ {\rm yr}^{-1}\ {\rm h_{75}^{-2}}$\  
for $\qno = 0.5$ and
$(180 - 910)\msun\ {\rm yr}^{-1}\ {\rm h_{75}^{-2}}$\ 
for $\qno = 0.02$, taking $\Psi=1.5$.
This is an order 
of magnitude higher than the star formation rates typically seen in
rapidly star forming galaxies observed in the optical\cite{pettini,steidel} 
and is similar to rates predicted from other recent submillimeter observations
(see, e.g., ref.~\citen{cimatti}).  Thus, our data indicate
that the upper end of the mass formation function is dominated by objects
that are radiating primarily in the submillimeter.

The relative fraction of the universal stellar energy which is
reradiated into the submillimeter and is present in these bright
submillimeter sources versus that which is radiated in the
rest frame ultraviolet is most simply quantified using the extragalactic
background light.  If we assume an Arp~220 spectrum for the submillimeter 
objects and a substantial fraction of light powered by star 
formation, then the sum of our two measured $>3\sigma$\ 850$\micron$ source 
fluxes, if located at $z=1$, would require reradiation of a
rest frame ultraviolet sky brightness of
$S_\nu = 9\times 10^{-25}\ \rm{ergs\ cm^{-2}\ s^{-1}\ Hz^{-1}\ deg^{-2}}$
to produce the bolometric submillimeter light.  [The redshift dependence of
$S_\nu$ is approximately~$((1+z)/2)^{-2.3}$].  In comparison,
the integrated flat spectrum population corresponding to the peak of 
the optically selected star
forming galaxies near $z=1$ gives a rest frame ultraviolet 
surface brightness\cite{songaila} of approximately
$4\times 10^{-25}\ \rm{ergs\ cm^{-2}\ s^{-1}\ Hz^{-1}\ deg^{-2}}$. 
While the submillimeter results are still quite uncertain due to 
the small number of sources,
these two values are quite comparable, with the submillimeter
light from the bright sources dominating if they lie at $z\approxlt 1.7$.
More accurate estimates will require the observations of many
SCUBA fields to the level described here.

In summary, our results suggest that high redshift star forming 
galaxies with the largest star formation rates radiate primarily in the 
submillimeter. Further support for this argument
has been obtained in very recent cluster observations\cite{blain2}.
Although these submillimeter sources with high star formation rates
are less common than are optically-selected galaxies,
the integrated light emitted by the submillimeter sources and by 
the optically-selected galaxies appears to be comparable within the
still considerable uncertainties in the surface densities.
The present results probe only the highest luminosity submillimeter
sources, and recent measurements of the integrated backgrounds indicate
that several times more energy could be present in fainter 
submillimeter sources\cite{puget,burigana,fixsen,schlegel,hauser}.

\acknowledgments
We are grateful to Remo Tilanus, Tim Jenness, Ian Smail, and William Vacca
for valuable interactions. We thank the two referees and the editor
for comments which led to substantial improvements in the paper.

\newpage

\centerline{\bf FIGURES}
{\parindent =0pc%\parskip =24pt

Figure~1\quad 850$\micron$ SCUBA maps of our two 
blank fields: (left) Lockman Hole and (right) SSA13. The top panels show
the total integrations, the middle panels show just the data from our 
May 1998 run, and the bottom panels show just the data from our March 1998 
run.  The vertical maps have their centers aligned.  The field centers
for the Lockman Hole are RA(2000) 10 33 55.4 and Dec(2000) 57 46 18
and for SSA13 are RA(2000) 13 12 25.6 and Dec(2000) 42 44 38. 
The diameter of each of the maps is $\sim 2.7$\ arcmin.
The exposure times for the maps are (a) 38.8\ hrs, (b) 8.1\ hrs, 
(c) 30.7\ hrs, (d) 51.0\ hrs, (e) 34.8\ hrs, and (f) 16.2\ hrs. 
The contours show levels of constant surface brightness.  In the central 
regions of the Lockman Hole maps the first contour level is 2.2\ mJy or 
$\sim 3\sigma$, and the second contour level is 4.4\ mJy or $\sim 6\sigma$.
In the central regions of the SSA13 maps the contour level is 2.2\ mJy or
$\sim 3\sigma$.  Any features right at the edges of the maps are insignificant 
at even the $1.5\sigma$ level due to the rapid increase in the noise levels 
at the very edge. Therefore, the only feature in the total maps other
than the two sources listed in Table~1 that we feel might be real is
the 2.5\ mJy feature approximately 45\ arcsec north of the identified source
in the SSA13 map.
The SCUBA observations were obtained by stepping the secondary
mirror through a 64-point jiggle pattern in Nasmyth coordinates while 
chopping at a frequency of 6.94 Hz and a chop throw of 45 arcsec in RA.
In order to cancel out the sky, the jiggle pattern was subdivided and
the target position switched between the signal and reference
beams using a repeating signal-reference-reference-signal scheme.
Because the chop throw is small the reference
negative images also appear on the array and hence can be restored,
increasing the effective exposure times on most of the field.  The primary 
source extraction was done using beam- and exposure-weighting to determine 
the signal and noise at each spatial position, and both positive and negative 
portions of the beam pattern were used.
Pointing checks were performed every hour using the blazars
0923+392 or 1308+326, and our data maps were calibrated using twice-nightly
beam maps of the calibration source CRL618 (March run) or thrice-nightly
beam maps of IRC+10216 (May run). The measurement of the calibration source
was found to vary by a maximum of $\sim 20$ per cent over the course of 
a shift.
``Skydips'', which measure the zenith atmospheric opacities at 450
and 850$\micron$, were done every few hours throughout each night,
and the sky opacity at 230\ GHz was monitored at all times to keep track 
of the stability of the sky.
The data were reduced in a standard way using the dedicated
SCUBA data reduction software SURF \cite{surf}.

Figure~2\quad (a) 3\,mJy/beam contour of the $\sim6\sigma$ source
detected in the 850$\micron$ Lockman hole map
displayed on a $K'$ image obtained with the QUick Infrared
Camera (QUIRC \cite{hodapp}) on the University of Hawaii
88-inch telescope on Mauna Kea during the period of 1995-1996. The units
on the axes are arcseconds, and the square box is the zoomed region 
shown in (b). (b) Zoomed $K'$ image of the source
(field center) obtained with the Near Infrared Camera 
(NIRC \cite{matthews}) on 
the Keck 10-m telescope in March 1998; $K'_{AB}=21.8$. The displayed
field is $\sim 45$\ arcsec on a side.
(c) Zoomed $B$ image obtained with the
University of Hawaii 88-inch telescope in March 1998; $B_{AB}=23.5$.
The displayed field is $\sim 45$\ arcsec on a side.
Magnitudes were measured in 6\,arcsec diameter apertures and put into
the AB magnitude system [${\rm AB} = -2.5\log f_\nu - 48.60$,
where $f_\nu$ is the flux in $\rm{ergs\ cm^{-2}\ s^{-1}\ Hz^{-1}}$].
}

%  TABLE 1
 
\begin{deluxetable}{c c c c c}
\tablecolumns{5}
\tablewidth{0pc}
\tablenum{1}
\tablecaption{850$\micron$ Sources}
\tablehead{
\multicolumn{4}{c}{} & \colhead{$1\sigma$ noise level} \\
\colhead{Field} & \colhead{RA(2000)} &
\colhead{Dec(2000)} & \colhead{Flux (mJy)} &
\colhead{at source position (mJy)} \\
}
\startdata
Lockman hole & 10 34 02.3 & 57 46 25.0 & 4.6 & 0.8 \\
SSA13 & 13 12 32.1 & 42 44 28.0 & 3.3 & 0.9 \\
\enddata
\tablecomments{The UT 1998 dates and
zenith 850$\micron$ sky optical depth ranges
for the March run were 17 (0.30 -- 0.31), 19 (0.42 -- 0.47),
20 (0.20 -- 0.35), 21 (0.13 -- 0.14), 22 (0.13 -- 0.30),
23 (0.13 -- 0.24), 24 (0.10 -- 0.11), 25 (0.10) and for the May
run were 16 (0.28 -- 0.41), 17 (0.27 -- 0.36), 18 (0.28 -- 0.38),
19 (0.45 -- 0.58), 22 (0.11 -- 0.14), 23 (0.11 -- 0.14),
24 (0.20 -- 0.23), and 25 (0.16 -- 0.25). During the March
run the airmass range for the Lockman Hole was
1.3 -- 2.4 and for SSA13, 1.1 -- 1.4.
For the May run the airmass range for the Lockman Hole
was 1.3 -- 1.5 and for SSA13, 1.1 -- 1.7.
}
\end{deluxetable}

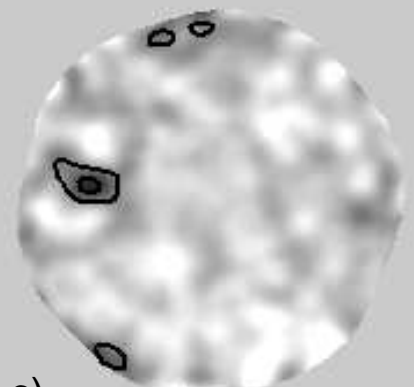
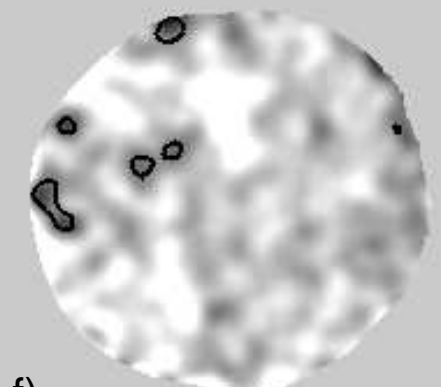
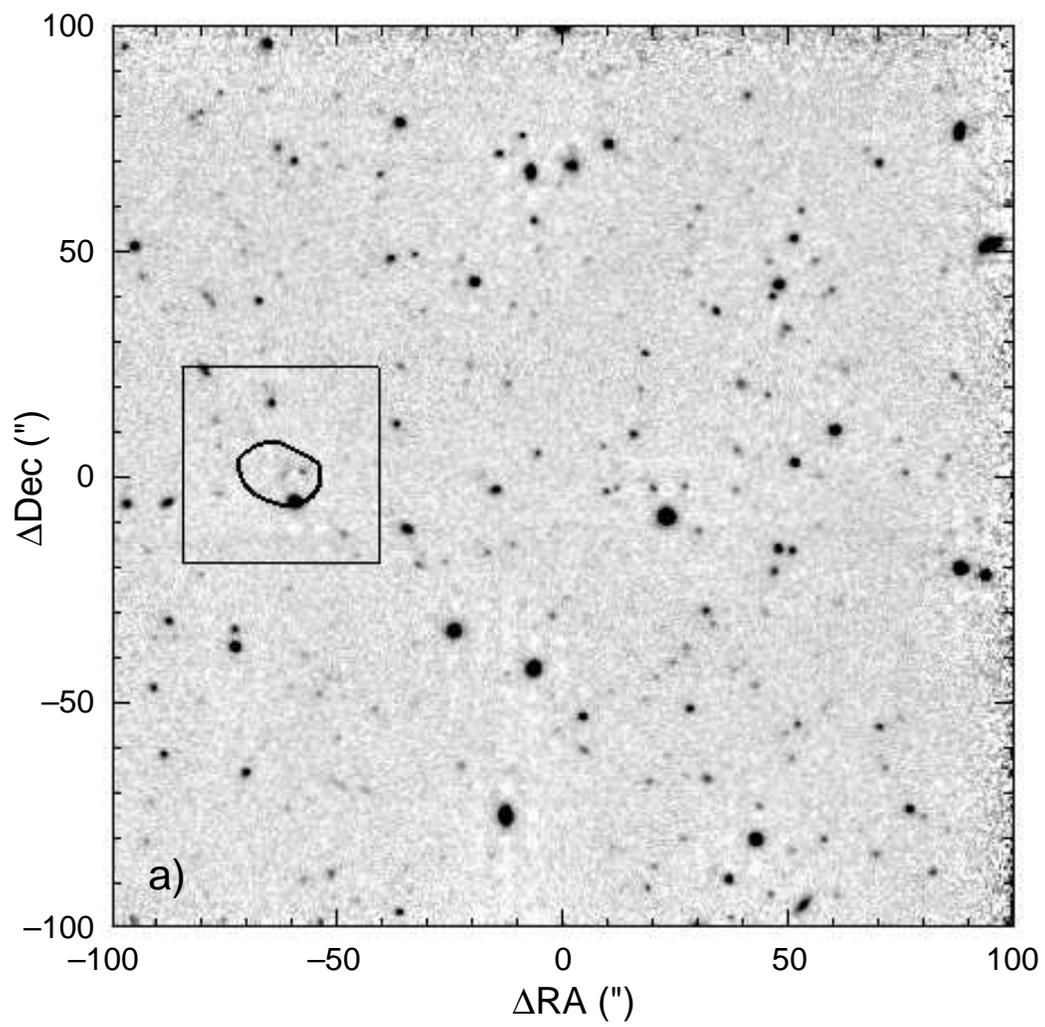
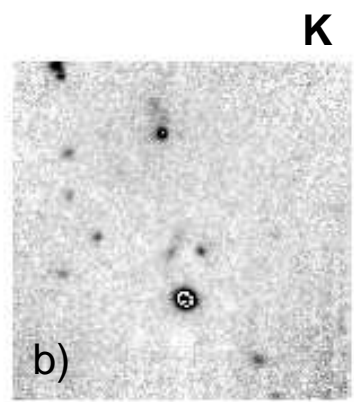
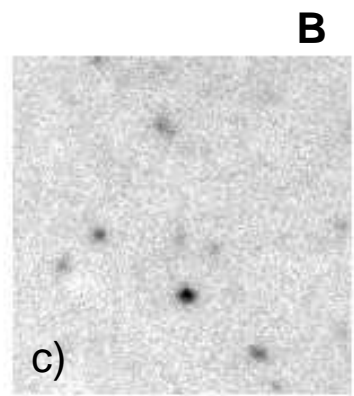
\end{document}

%% file: basic.tex
\def\aanat#1{{\it Astr. Astrophys.\/} {\bf #1}}

\def\ajnat#1{{\it Ast. J.\/} {\bf #1}}
\def\apjnat#1{{\it Astrophys. J.\/} {\bf #1}}

\def\araanat#1{{\it A. Rev. Ast. Astrophys.\/} {\bf #1}}
\def\mnnat#1{{\it Mon. Not. R. Astr. Soc.\/} {\bf #1}}
\def\paspnat#1{{\it Pub. Ast. Soc. Pac.\/} {\bf #1}}
\def\natnat#1{{\it Nature\/} {\bf #1}}
\def\nanat#1{{\it New Astronomy\/} {\bf #1}}

\def\ten#1{\ifmmode 10^{#1}\else $10^{#1}$\fi}

\def\approxgt{\lower.2em\hbox{$\buildrel > \over \sim$}}
\def\approxlt{\lower.2em\hbox{$\buildrel < \over \sim$}}
\def\uline#1{$\underline{\hbox{\kern #1in}}$ }
\def\sqr#1#2{{\vcenter{\hrule height .#2pt
          \hbox{\vrule width .#2pt height#1pt \kern#1pt
                  \vrule width .#2pt}
          \hrule height .#2pt}}}

\def\88in{$88^{\prime\prime}$}
\def\24in{$24^{\prime\prime}$}

\def\msun{~\rm M_{\odot}}
\def\lsun{~\rm L_{\odot}}

\def\cm2{~\rm cm^{-2}}

\def\chisq{\ifmmode {\chi^2}\else {$\chi^2$}\fi}

\def\qno{q_0}